# ASTRONOMICAL OBSERVATORY PUBLICATIONS: INFORMATION EXCHANGE BEFORE THE INTERNET AGE

## Ole Ellegaard and Bertil F. Dorch

*University Library of Southern Denmark, Campusvej 55,*
*5230 Odense M, Denmark.*
E-mails: oleell@bib.sdu.dk; bfd@bib.sdu.dk

**Abstract:** For decades, perhaps even centuries, the exchange of publications between observatories was the most important source of information on new astronomical results, either in the form of observational data or new scientific theories. In particular, small observatories or institutions used this method. The exchange of physical material between observatories has now been replaced by the exchange of information via the Internet. Yet much of the ancient material has never been digitized and can only be found in the few existing collections of observatory publications. A recent donation of such a collection from the University of Copenhagen to our own library at the University of Southern Denmark has led us to investigate the uniqueness of such collections: Which observatories and publications are represented in the collections that still exist today? We also examine the availability of the material in the collections.

## 1 INTRODUCTION

In the past, astronomical observatories exchanged the results of their scientific work with other observatories and societies around the world. This scientific exchange probably took place since observational astronomy arose, and at least on a regular basis from the beginning of the nineteenth century. It was for a long time considered the 'proper' way to publish observations (Jaschek, 1989: 152). In fact, much of astronomy is based on observations with telescopes, and every observation is unique and should in principle be kept as a historical record (de Narbonne, 1989). Eichhorn, et al. (2003: 148) say:

> Throughout the 19th century and well into the 20th century, many, if not most, of the important articles in astronomy were published in observatory publications (OPs).

Initially, several almanacs, calenders or 'Jahrbuchs' had already been published, which also contained examples of astronomical observations (Holl and Vargha, 2003). Observatories could also present forthcoming observations and outlines of their scientific programs in these publications (ibid.). The literature varies from collection to collection at the various libraries and consists of tables of observations, bulletins, annual reports, circulars, newsletters, reprints, etc.

This distribution and use of observatory publications (OPs) occurred on a smaller scale before and continued at the same time as the introduction of ordinary astronomical journals. Examples of the latter are *Monatliche Correspondenz zur beförderung der Erd und Himmelskunde* (1801) and *Astronomische Nachrichten* (1821). In fact, Corbin and Coletti (1995)

wrote:

> A large part of the important research in astronomy during the late 19th and early 20th centuries was reported in great detail in exhaustive articles in the main publication series of observatories throughout the world. Although journals were also important during this period, the reports of these research projects containing descriptions of instruments, detailed tables of findings and explanations of the course of the research that were too long to be published in journals.

The first astronomical journals and the regular distribution of OPs began around the late period of the European Enlightenment, but technological development and the possibility of public funding may also have played a role in this development (Burns, 2003).

OPs are scientific publications unique to astronomy and have proven to be crucial for many observatories, especially in smaller locations that could not support a subscription-based collection of the major astronomical journals. As stated by Kaminska, (1989: 137):

> Any free publications obtained by my library constitute an extremely valuable source of information about new research and discoveries all over the world.

The emergence of OPs could be seen as a continuation of a tradition of sharing observatory data among the few observatories that existed around the beginning of the nineteenth century. It was rooted in strong, individual institutions that hosted the observatories, but the exchange was particularly useful for institutions with small budgets.

In many cases, OPs lead to a faster exchange of new ideas and observations than did traditional publication in journals. The material was often in a form that did not fit into regular magazines or books. Furthermore, more data from observations could be presented in OPs than was usually the case for ordinary papers. On the one hand, Crabtree (2018), who recently analysed the impact and performance of observatories, did not distinguish between the two types of publications. On the other hand, the possibility of peer review in regular journals is mentioned as an advantage compared to OPs, although automatically, universal peer review only became common in the late twentieth century (Holl, 2004). Therefore, peer review played no role in the early stage of publishing astronomical results which, for example, was similar to the field of physics. The lack of periodicity as well as the increased cost compared to publishing in regular journals ultimately proved to be detrimental to the OPs (Jaschek, 1989: 152). The paper by Holl (2004) provides a comprehensive overview of the publishing history of astronomy.

Today, of course, astronomical observations and ideas are exchanged more quickly via the observatories' own web pages, institutional repositories or astro-ph in arXiv.org, although relevant information related to the data management of collected astronomical data can sometimes be difficult to find in regular publications. Publication in electronic journals, archives from parent institutions or via external institutions is used for ordinary papers. The latter also allows the inclusion of series of observations as additional material.

The field of astronomy has a tradition of Open Access to scientific results—earlier through exchange of preprints and OPs but today primarily web based—which has been to the benefit of society as a whole and especially in countries with smaller budgets for research. Publication takes place either as 'golden' open access, when the authors pay the journal a publication fee or as 'green' open access, when the publication is deposited in archives such as astro-ph in arXiv.org (Ginsparg, 2011). The latter unfortunately often apply a quarantine period if the paper is published in an ordinary journal, the so-called 'delayed green open access', although core journals like *Astronomy & Astrophysics* (*A&A*) have no delay. We can regard the exchange of observatory publications as a predecessor to modern forms of Open Access in astronomy, continuing a longstanding astronomical culture of openness.

Material in the old paper-based observatory publications is in many cases not particularly relevant to modern astronomy, although some results may still be valid and useable (Coletti,

2003). Examples may be data on variable stars, sunspots or observations of comets. To a large extent, the material is of great historical interest in documenting the timeline for achieving scientific results, and astronomical communities and their working methods (Corbin and Coletti, 1995). With limited space in regular journals, more detailed observational data are often found in OPs.

In many cases, observatory's publications have not survived the passage of time or have been discarded because the research field or the observatory at the institution has been closed. The books and pamphlets are often without hard bindings, and in many cases were printed on acidic paper, so they discolour and become fragile. As a result, more or less complete collections of OPs are rarely found today, and as early as the 1990s, plans for digital preservation of complete publication series from observatories around the world already existed (ibid.). This is in contrast to the case of paper-based astronomical journals, which were almost fully digitized before 2003 (Eichhorn, et al., 2003), although it turned out that this task was somewhat more difficult to achieve than anticipated.

Accessing publications in either physical or electronic format is one thing, but first you have to identify the right publication. Until the year 2000, the bibliography 'Astronomy and Astrophysics Abstract' produced by the Astronomisches Rechen Institute in Heidelberg, Germany, registered observatory publications and published a list of the indexed observatories and institutes. Much of the information derived from the Astronomy and Astrophysics Abstract is also available on-line in the physical and astronomical database INSPEC. Today, the large SAO/NASA Astrophysical Data System (abbr. ADS), launched in 1993, has become the dominant bibliographic tool to identify astronomical literature, including observatory publications.

Our main focus in this paper is not to present a historical account of OPs in general, as this has already been done in a review paper by Holl and Varga (2003). Instead, a recent donation of a large collection of observatory publications from the Astronomical Observatory at the University of Copenhagen, Denmark, to the University Library at the University of Southern Denmark has spurred our interest in the following matters: How comprehensive and unique is our 'new' collection? Collections of OPs are said to be rare, so to what extent can we find the material elsewhere in other institutions and ultimately in digitized format?

Is it possible to estimate how many large collections with a significant number of OPs that still exist today? How complete are the collections in terms of observatories presented as well as individual titles? It is also essential that the collections are recorded, that they can be referenced and are accessible for use by the public. Implementation of the latter has already been discussed in some detail by Corbin and Coletti (1995), but our current study shows that the process of full registration and digitization is far from complete. Due to the often-fragile condition of the material and the risk of being involved in scrapping procedures, it may ultimately jeopardize the goal of preserving these historical records.

Table 1: Various 'large' collections of observatory publications.

| Nation | US State | City | Library |
|--------|----------|------|---------|
| Denmark | | Odense | University Library of Southern Denmark |
| England | | Cambridge | Institute of Astronomy Library, Cambridge University |
| Poland | | Warzaw | Library of Warzaw University Observatory |
| Scotland | | Edinburgh | Royal Observatory Library |
| USA | Arizona | Tucson | NOIRLab Library |
| | New York | Albany | Dudley Observatory Library |
| | Massachusets | Cambridge | Wolbach Library, Harvard College Observatory |
| | [District of Columbia] | Washington | James Melville Gilliss Library, U.S. Naval Observatory |
| | Wisconsin | Madison | Woodman Library, University of Wisconsin |

## 2 COLLECTIONS OF ASTRONOMICAL OBSERVATORY PUBLICATIONS

An informal search on the Internet revealed that large collections of OPs seem to be rare. Notable examples are listed in Table 1. It is also evident that smaller collections are found in many other observatory libraries worldwide. United States appears to have the largest number of collections. In fact, it is claimed that the combined collections of the Harvard College Observatory (Wolbach) and the Naval Observatory in Washington are the most complete in the United States today (see: library.harvard.edu/collections/astronomical-institutions-ai-collection). The Wolbach Library is a key library for the preservation and identification of OPs. The Hollis catalogue in the Library can be searched for its own OPs as well as those at the Naval Observatory. Much of the material has been microfilmed and is available in full text via the 'Harvard Astronomy Preservation Microfilm Project': https://library.cfa.harvard.edu/microfilmproject (Coletti, 2002). A Hollis search leads to 250 hits that include an observatory, though not all are unique.

The ADS, Astrophysical Data System Abstract Service is also collaborating with the Wolbach Library at the Smithsonian Centre for Astrophysics on the digital preservation of OPs based on previous microfilm scans. The transfer of the documents has proved technically difficult with the risk of loss of information (Col-etti and Corbin, 2003). The 'Historical scans of Ops' in ADS include a list of the observatories involved (adsabs.harvard.edu/historical.html). In contrast, there is no complete list of observatories in the Wolbach Library or via the Hollis catalogue. The Observatory's records must be found individually. In fact, a list of observatories in the collection is useful for gaining an overview.

A distinct problem with regard to searching in the older astronomical literature, including observatory publications, is the lack of publication metadata (i.e. structured information used to search, describe or manage documents). Examples of this could be standardized keywords, indexes or abstracts (Corbin and Coletti, 1995). Initially, the Harvard Library attempted to address this voluntarily through public use of metadata capture programs (Eichhorn, et al., 2003).

Complete observatory lists can be found in the case of the collections from the Woodman Library, the Dudley Observatory and our own collection from the Astronomical Observatory at the University of Copenhagen. The Woodman Library has a regular 'Observatory publication finding aid' as an Excel file with complete entries for each publication.

The other two observatories have less accessible catalogues—the Dudley Observatory in the form of scanned catalogue cards, and in the case of the Danish collection, the same type of card is still in printed format (sometimes with handwritten records that cannot be read by OCR technology). The bibliographic entries include the name of the observatory, the city of the observatory, the title of the publication, language, year and some holding information. The city, where the observatory is located, is the most important item in the catalogue. A number of major cities hosted several different observatories. In some cases, the observatories have changed over time, and some observatories have been closed.

To identify the uniqueness of the various collections, we counted the number of observatories included from each library. Table 2 shows the data from the three aforementioned libraries supplemented by the list of observatories indexed in 'Historical scans of Ops' in ADS. In all, 945 towns or observatories are represented in this table, of which 436 are different (46.1%). We found that there are 150 observatories present in the Denmark, Woodman and Dudley Library collections, while 80 observatories are present in all four collections.

Table 2: Observatories/towns included in different library collections.

| Library | Library of Southern Denmark | Woodman Library | Dudley Observatory Library | ADS Historical Scans |
|---|---|---|---|---|
| Towns/Observatories present in the collection | 319 | 264 | 252 | 110 |
| Unique Towns/Observatories | 79 | 25 | 61 | 4 |

Table 3: Some examples of the different holdings in the libraries that are available for the same observatories. In the case of individual titles, the Hollis catalogue at the Harvard Library (hollis.harvard.edu) provides the most comprehensive search and identification of OPs at different locations.

| Observatory | Library of Southern Denmark | Woodman Library University of Wisconsin (Madison) | Dudley Observatory Library | ADS Historical Scans |
|---|---|---|---|---|
| Besancon, France | *Annals* 1934−1936, 1939−1972 *Bulletin Astronomique* 1986−1995 *Bulletin Chronometrique* 1889−1902,1904−1925 | *Annals* v. 3, no. 1−3, 1934−1944 v. 4, no. 1−5, 1944−1955 | *Annals* 1936−1950 | *Annals* 1934−1964 |
| Lowell, Flagstaff, Arizona. USA | *Annals* 1898−1905 *Bulletins* 1911−1983 *Memoirs* 1915 | *Annals* v. 1−3, 1898−1905 | *Annals* 1898−1905 *Bulletins* 1903−1978 First catalogue (reprint) 1898 *Memoirs* 1915 | *Annals* v. 1−3, 1898−1905 |
| Buenos Aires, Argentina | IAFE, Serie *Publicaciones de registros. Tirada aparte* 1−10, 1971−1975, 14 1977 IAFE, 1−32, 45−47, 49−57, 59, 1971−1982 | Missing? | Not present | Not present |
| Jodrell Bank Manchester, England | *Annals* 1−6, 1952−1960 *Reprints* Vol. 1−8 *Astronomical Contributions* 2−228, 230−234, 236−250, 1950−1971 | *Annals* 1952−1960 *Reprints* v. 1−8, 1946−1969 *Astronomical Contributions* (Series 3) no 1−250, 1952−1971 | Not present | Not present |
| Kodaikánal Bangalore, India | *Newsletter* 1−8, 1986−1894 *Bulletin* 1−11, 1908−1970 *Series A,B,C*, 1967−1997 *Annual Report* 1898−1991 *Memoires* Vol. 1−2, 1909−1917 *Reprints* 1957−1975 *VBT News* 1−9, 1989−1991 | *Annual Report* 1987−1989 | *Bulletin* 1908−1969 *Memoire* 1909− | *Bulletin* Vol. 1−4, 1908−. |

The Danish collection includes the largest number of different cities/observatories and also the largest number that can only be found here. The Dudley Library has an almost identical number of 'unique' observatories in its collection. The observatories present in 'Historical scans' in ADS represent approximately 25% of the total number of different observatories in all four collections and is currently far from complete. Furthermore, the records for each observatory are very different from library to library, both in terms of titles and range of years present. Examples of this are shown in Table 3.

As a consequence, any search for a specific publication in a collection means that one must identify whether the city/observatory is present, then the title and the document's number and year. If you are lucky, the document has been digitized, but in many cases, you have to try to find the original document or obtain a photocopy of the same. The latter is often preferred due to the fragile condition of the material. Observatory publications were regularly printed on cheap, acidic paper.

## 3 COLLECTION OF THE DANISH ASTRONOMICAL OBSERVATORY

The collection of the University of Southern Denmark's Library originates from the now discontinued astronomical library at the Niels Bohr Institute in Copenhagen, Denmark. Detailed information about the origin of the collection and the Observatory Library at Østervold, Copenhagen, has recently been published in this journal (Dorch and Petersen, 2021). Our collection contains materials from 319 different observatories around the world. The material has been collected for more than a hundred years, and the oldest material dates back many centuries. At present, the collection is only reg-

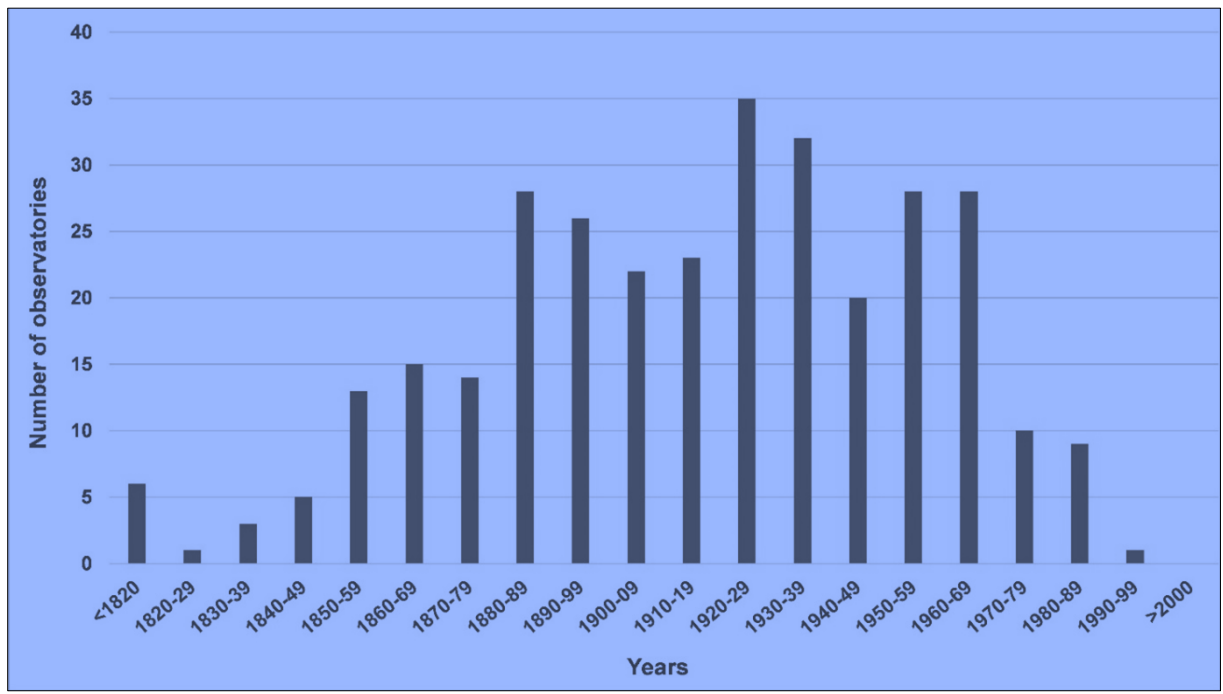

Figure 1: The Danish collection, showing the number of observatories with the first registered publication in a given decade.

istered on catalogue cards, but plans for digital registration are underway. Figure 1 shows the extent of the years included. The number of observatories with the first registered publication in a given decade is depicted. It is clear that the exchange of observatory publications took place as early as the beginning of the nineteenth century and reached its peak in the decades before World War II. This event led to a decline in activity in the following period, but the exchange was revived and continued until the beginning of the electronic and digital era in the 1970s leading to a complete change in the way information was distributed among the astronomical libraries. This 'information revolution' not only meant a faster exchange of astronomical observations and ideas, but the storage of material (or data) became very different. In the wake of this, problems with rapidly obsolete information technology and disorganized data gave rise to new challenges

Shown in Figures 2 and 3 are a few examples of the material found in the collection from Copenhagen University.

## 4 DISCUSSION

Observatory publications in paper format have in the past proved to be a vital and inexpensive form of exchanging scientific results. Although not normally peer-reviewed, the quality of the material is guaranteed by the observatory's director and reputation. Despite the hundreds of observatories around the world that have been involved in these exchanges, our analysis shows that only a few large collections seem to exist today. These collections are far from complete either in terms of the number of observatories or their publications. As a consequence, Holl and Vargha (2003) suggested that old OPs should be made available via the Internet. This has been achieved in part at the Wolbach Library through the 'Harvard Astronomy Preservation Microfilm Project' and followed by the digitization of the material in ADS. Our analysis also shows that a maximum of 25% of the observatories are actually represented in the data from the 'historical scans' at ADS. In terms of bibliographic or 'meta' data on OPs, the 'observatory-publications-finding-aid' at the Woodman Library has proven to be invaluable and to our knowledge the most comprehensive of its kind. Our own collection has never been digitized and bibliographic information is only available on cards. On the other hand, with 79 unique observatories and individual publications in the collection, at least a partial digitization and registration in a database would be an advantage to do.

However, even if the entire record of OPs could be digitized and made openly available, in the extreme long term, several technical and cultural issues remain relating to the general sustainability of digital material (cf., for example, the discussion in Dorch et al. (2019) in the context of digitizing astronomical heritage).

Information technology is now so advanced that a modern, free and web-based form of observatory publications can re-emerge, as al-ready suggested by Holl and Varg, (2003). The traditional form of publication in peer-reviewed journals has in many cases been too slow for e.g., the dynamic, observational astronomy.

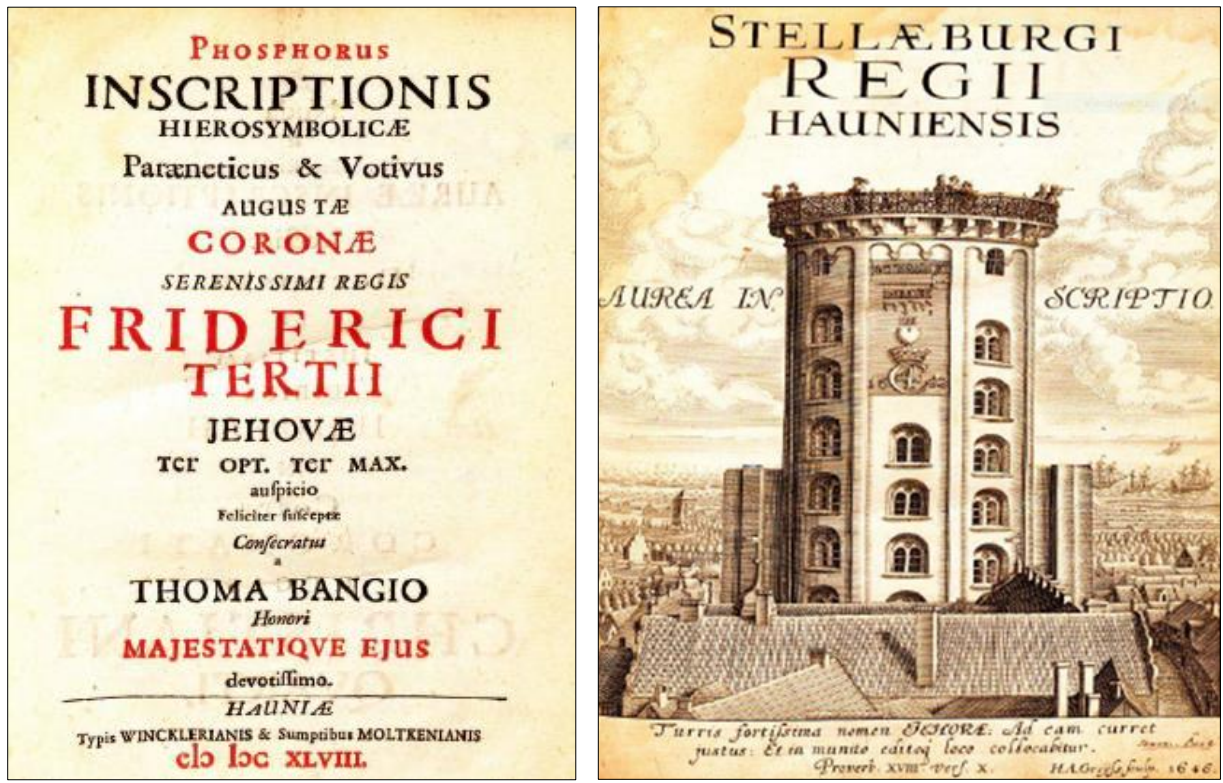

Figure 2: The oldest OP in the Danish collection: *Phosphorus INSCRIPTIONIS*. It depicts the newly built observational tower in the centre of the capital, Copenhagen (after Bangio, 1648).

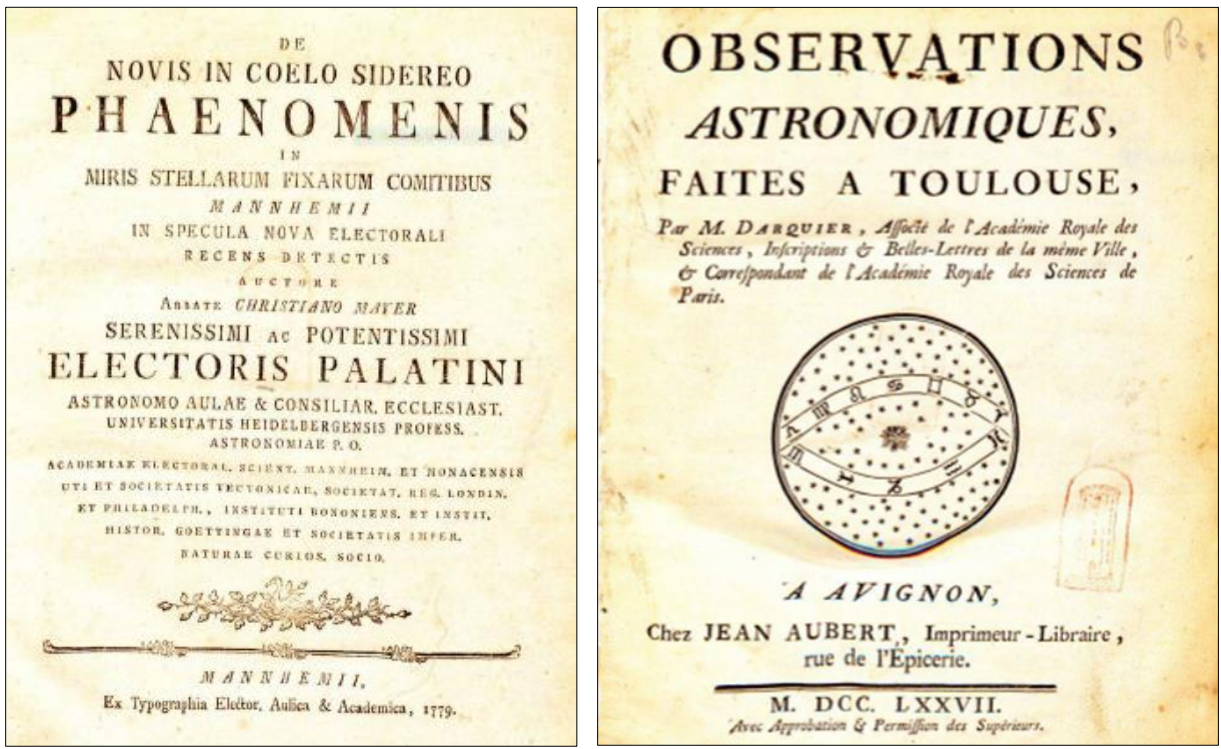

Figure 3: Two examples of early OPs. The first is from the observatory in Mannheim on star companions (after Mayer, 1779) and the second from Toulouse on astronomical observations at the observatory (after Darquier, 1777).

This is even worse if the publications are hidden behind a payment wall. Of course, there are open access repositories, such as arXiv.org, but online observatory platforms are more flexible and can put the results obtained in the right context. It is also clear that there are other challenges with maintaining and storing the publications. The printed versions of OPs, although sometimes in a fragile state, are still with us.

As we have noted, a number of OPs are only found in some institutions, and not all material is digitized. This increases the risk of material being lost due to institutional changes or cuts to a library's budget. A good example is related to our own collection, as described by Dorch and Petersen (2021). The question is whether in the distant future we will have access to the material available today. It is also important that the information is searchable. This task is solved in an excellent way, in the case of traditional journal articles, by the major bibliographies such as SAO/NASA ADS or the Web of Science. When dealing with information on web platforms, we rely heavily on search engines like Google that are effective but not very transparent in terms of the weighting of search results, and the search options are also very limited. As a result, in terms of astronom-ical information, you need to choose between well-organized bibliographies that lead to journal papers or less organized, diverse material on various web pages.

## 5 CONCLUSION

We have examined the presence of paper-based astronomical observatory publications in a number of libraries. Few seem to have large collections of OPs. Despite the fact that a number of observatories are represented in several libraries, we also observe a number of unique observatories that are only present in one library. We found that this was especially the case for our own collection from the University of Copenhagen in Denmark. The individual stocks and years included also vary from library to library. A small part of the material has actually been digitized (representing about 25% of all observatories) and is included in the list of 'digital scans' at SAO/NASA. A comprehensive search of ancient astronomical observations and material in general should therefore include a number of libraries that still have preserved large collections of observatory publications.

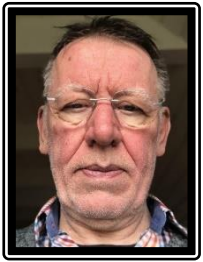

**Dr Ole Ellegaard** received his PhD in Physics from Risø National Laboratory, Denmark in 1986. The field was experimental physics and the study of the interaction of keV electrons and ions with condensed gases, which is relevant in the context of planetary science. After graduating, he worked as a research assistant and science teacher.

From 1988, he was employed as a research librarian in physics and chemistry at the University of Southern Denmark. Several papers were published in the field of particle surface interactions and laser ablation, but lately his interests have been in the fields of science policy, history of science, and especially bibliometrics with some highly cited articles in this field. Currently, Ellegaard held an emeritus position at the University Library of Southern Denmark.

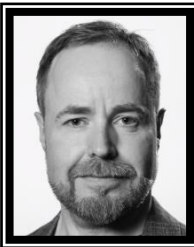

**Dr Bertil F. Dorch** was born in Denmark in 1971, and is an Associate Professor in the Department of Physics, Chemistry and Pharmacy at The University of Southern Denmark (SDU), and since 2013 Director of the Research and University Library at SDU. He received his PhD in Physics from The Niels Bohr Institute at the University of Copenhagen in 1998, where he also obtained his Master's, in Astronomy, in 1995. He previously held research positions at the Royal Swedish Academy of Science and at the University of Copenhagen.

Apart from teaching, Dorch participates in astronomy outreach, and is active in public and research political debate. Between 2014 and 2020 on three occasions he was President of the Danish Research Library Association. He is a member of the International Astronomical Union, and currently serves as a board member of various national and international organizations.